\documentclass[12pt]{article}
\usepackage{amsmath}
\usepackage{amssymb}
\usepackage{amsfonts}
\usepackage{graphics}

\newcommand{\field}[1]{\mathbb{#1}}
\newcommand{\R}{\field{R}}
\DeclareMathOperator{\arctanh}{arctanh}

\title{
Generalized nonlinear oscillators with quasi-harmonic behaviour: classical solutions}
\author{C. Quesne\thanks{Electronic mail: cquesne@ulb.ac.be}\\ 
{\small\sl Physique Nucl\'eaire Th\'eorique et Physique Math\'ematique,  Universit\'e Libre de Bruxelles,} \\ 
{\small\sl Campus de la Plaine CP229, Boulevard~du Triomphe, B-1050 Brussels, Belgium}}
\date{ }
\begin{document}
\baselineskip=22pt plus 1pt minus 1pt
%%%%%%%%%%%%%%%%%%%%%%%%%%%%%%%%%%%%%%%%%%%%%%%%%%%%%%%%%%
\maketitle
\begin{abstract}
The classical nonlinear oscillator, proposed by Mathews and Lakshmanan in 1974 and including a position-dependent mass in the kinetic energy term, is generalized in two different ways by adding an extra term to the potential. The solutions of the Euler-Lagrange equation are shown to exhibit richer behaviour patterns than those of the original nonlinear oscillator.
\end{abstract}

\noindent
Running title: Generalized nonlinear oscillators 


%
%========================================================================
%
\newpage
\section{INTRODUCTION}

In 1974, Mathews and Lakshmanan \cite{mathews, lakshmanan} introduced a classical nonlinear oscillator as a one-dimensional analogue of some quantum field theoretical models. This system was described by a Lagrangian
\begin{equation}
  L = \frac{1}{2} \frac{1}{1+ \lambda x^2} (\dot{x}^2 - \alpha^2 x^2),  \label{eq:L}
\end{equation}
which was a $\lambda$-dependent deformation of that of the standard harmonic oscillator. Apart from the nonlinearity of the oscillator potential, it presented the interesting feature of containing a position-dependent mass term in the kinetic energy. As a result, the equation of motion
\begin{equation}
  (1 + \lambda x^2) \ddot{x} - \lambda x \dot{x}^2 + \alpha^2 x = 0  \label{eq:E-L}
\end{equation}
had some solutions with a simple quasi-harmonic form, with the restriction that the frequency was amplitude dependent.\par
%
%--------------------------------------------------------------------------------------------------------------------------
%
A lot of studies have been devoted to this nonlinear oscillator, to its quantum version, and to some generalizations thereof.\par
%
%-----------------------------------------------------------------------------------------------------------------------------
%
It was pointed out \cite{carinena04a} that using a position-dependent mass was equivalent to changing the measure on the line, which is a general property of position-dependent mass systems \cite{cq}. This allowed the quantization of the Hamiltonian function corresponding to (\ref{eq:L}) without having to solve the usual ordering ambiguity problem of the momentum and mass operators \cite{vonroos}. The $\lambda$-dependent eigenvalues and eigenfunctions of the Hamiltonian were obtained in explicit form \cite{carinena04a, carinena07a} by taking advantage of the factorization and deformed shape invariance methods (see also Ref.~\cite{bagchi}). The classical polynomials arising in the bound-state wavefunctions were identified later on \cite{midya, schulze12}. Some generalized quantum potentials were also proposed  and the corresponding Schr\"odinger equations were solved \cite{midya, schulze13, ranada}.\par
%
%-----------------------------------------------------------------------------------------------------------------------------
%
On the other hand, a two-dimensional (and more generally $n$-dimensional) version of the one-dimensional system (\ref{eq:L}) was proposed and studied both at the classical \cite{carinena04b} and at the quantum \cite{carinena07b, carinena07c} levels. In both cases, the solutions were given, the system was shown to be superseparable and superintegrable, and the model was related to that of the harmonic oscillator in spaces of constant curvature $\kappa = -\lambda$. It was shown, in particular, that in classical mechanics all the bounded motions were quasiperiodic oscillations and that the unbounded (scattering) motions were represented by hyperbolic functions, as in one dimension.\par
%
%-------------------------------------------------------------------------------------------------------------
%
The purpose of the present paper is to reconsider the one-dimensional model with the same kinetic energy term as in (\ref{eq:L}), but with two more general nonlinear oscillator potentials.\par
%
%-----------------------------------------------------------------------------------------------------------------
%
In Secs.~II and III, the generalized potentials are presented, the corresponding Euler-Lagrange equations are solved and their solutions are compared with those of (\ref{eq:E-L}). Finally, Sec.~IV contains the conclusion.\par
%
%======================================================================
%
\section{FIRST GENERALIZED NONLINEAR OSCILLATOR}

\setcounter{equation}{0}

\subsection{The potential}

Let us consider the potential
\begin{equation}
  V(x) = \frac{1}{2} \frac{\alpha^2 x^2 - 2\beta x}{1 + \lambda x^2},
\end{equation}
depending on an additional parameter $\beta$ and reducing to the Mathews and Lakshmanan potential for $\beta$ going to zero. We will distinguish between two cases according to the sign of the deformation parameter $\lambda$:
\begin{eqnarray}
  & ({\rm i}) \; &  \lambda < 0, \quad - \frac{1}{\sqrt{|\lambda|}} < x < \frac{1}{\sqrt{|\lambda|}}, \quad 0 < \beta <
         \frac{\alpha^2}{2\sqrt{|\lambda|}};  \\
  & ({\rm ii}) \; &  \lambda > 0, \quad -\infty < x < \infty, \quad  \beta > 0.
\end{eqnarray}
\par
%
%----------------------------------------------------------------------------------------------------------------------
%
In the first case, the potential goes to $+\infty$ at both ends of the interval $\left(- 1/\sqrt{|\lambda|}, 1/\sqrt{|\lambda|}\right)$. Between its two zeros at $x=0$ and $x= 2\beta/\alpha^2$, it has a minimum
\begin{equation}
  V_{\rm min} = V(x_{\rm min}) = - \frac{\beta}{2} x_{\rm min} \qquad \mbox{at} \qquad x_{\rm min} = 
  \frac{\alpha^2 - \sqrt{\alpha^4 - 4|\lambda|\beta^2}}{2|\lambda|\beta}.
\end{equation}
In the second case, the potential, which goes to $\alpha^2/(2\lambda)$ for $x \to \pm\infty$, has still two zeros at $x=0$ and $x=2\beta/\alpha^2$, between which it has a minimum 
\begin{equation}
  V_{\rm min} = V(x_{\rm min}) = - \frac{\beta}{2} x_{\rm min} \qquad \mbox{at} \qquad x_{\rm min} = 
  \frac{-\alpha^2 + \sqrt{\alpha^4 + 4\lambda\beta^2}}{2\lambda\beta}.
\end{equation}
However, this time, for $x<0$ there also occurs a maximum
\begin{equation}
  V_{\rm max} = V(x_{\rm max}) = - \frac{\beta}{2} x_{\rm max} \qquad \mbox{at} \qquad x_{\rm max} = 
  \frac{-\alpha^2 - \sqrt{\alpha^4 + 4\lambda\beta^2}}{2\lambda\beta}.
\end{equation}
\par
%
%--------------------------------------------------------------------------------------------------------------------------------
%
\begin{figure}[h]
\begin{center}
\includegraphics{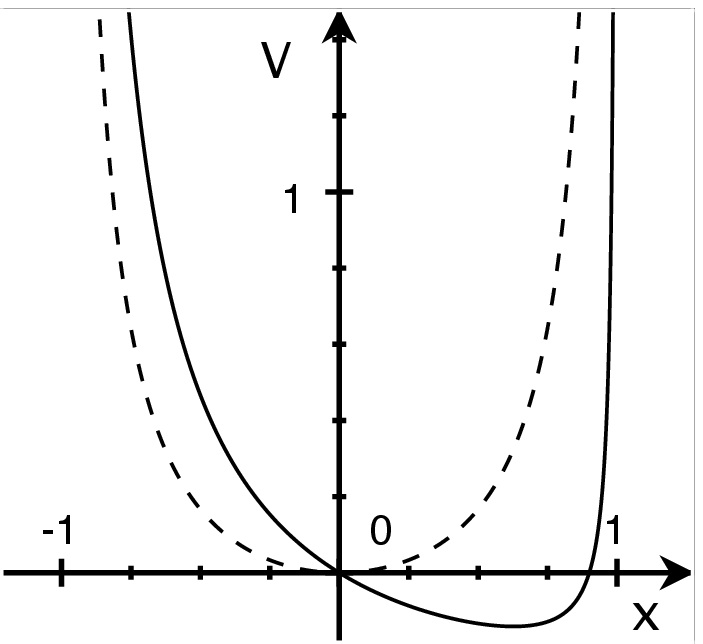}
\caption{Plot of $V(x)= \tfrac{1}{2} (\alpha^2 x^2 - 2\beta x)/(1 + \lambda x^2)$, $\alpha = -\lambda = 1$, as a function of $x$, for $\beta = 0.45$ (solid line) and $\beta = 0$ (dashed line).}
\end{center}
\end{figure}
\par
%
%-------------------------------------------------------------------------------------------------------------------------------
%
\begin{figure}[h]
\begin{center}
\includegraphics{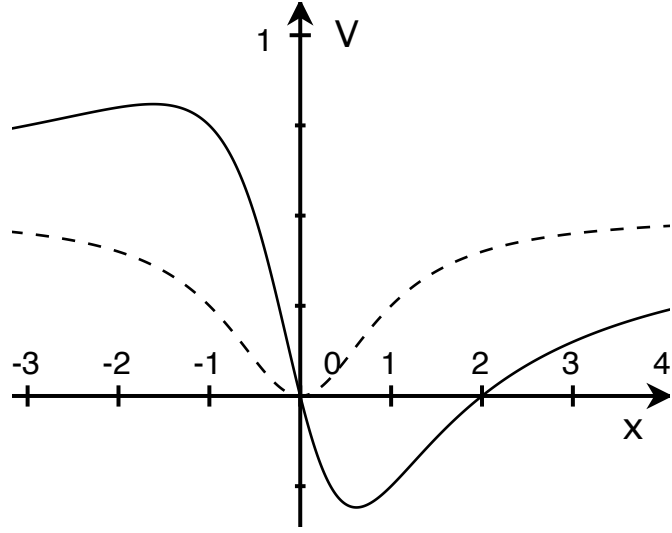}
\caption{Plot of $V(x)= \tfrac{1}{2} (\alpha^2 x^2 - 2\beta x)/(1 + \lambda x^2)$, $\alpha = \lambda = 1$, as a function of $x$, for $\beta = 1$ (solid line) and $\beta = 0$ (dashed line).}
\end{center}
\end{figure}
\par
%
%-------------------------------------------------------------------------------------------------------------------------------
%
In Figs.~1 and 2, some examples are plotted and compared with the $\beta=0$ case. Apart from the symmetry breaking with respect to parity that occurs when $\beta$ becomes nonvanishing, no substantial change is noted  in the $\lambda < 0$ case since the potential remains a well with boundless walls at $x = \pm 1/\sqrt{|\lambda|}$, thus allowing only bounded trajectories. In contrast, in the $\lambda > 0$ case, the advent of a maximum for $x < 0$ produces some changes: if for small energies, the trajectories remain bounded, for $E > \alpha^2/(2\lambda)$ there is now a distinction between the unbounded trajectories below or above the maximum $V_{\rm max}$, as well as two limiting unbounded motions at $E = \alpha^2/(2\lambda)$ and $E = V_{\rm max}$ instead of a single one at $E = \alpha^2/(2\lambda)$.\par
%
%++++++++++++++++++++++++++++++++++++++++++++++++++++++++++++++++++++
%
\subsection{Euler-Lagrange equation}

It is straightforward to see that the Euler-Lagrange equation (\ref{eq:E-L}) now becomes
\begin{equation}
  (1 + \lambda x^2) \ddot{x} - \lambda x \dot{x}^2 + \alpha^2 x - \beta (1 - \lambda x^2) = 0.
\end{equation}
To solve this equation, we proceed in two steps. First, let us set $\dot{x} = p(x)$, so that we obtain a first-order equation for $p^2$,
\begin{equation}
  (1 + \lambda x^2) \frac{dp^2}{dx} - 2\lambda x p^2 + 2\alpha^2 x - 2\beta (1 - \lambda x^2) = 0,
\end{equation}
whose general solution is given by
\begin{equation}
  p^2(x) = C (1 + \lambda x^2) + \frac{\alpha^2}{\lambda} + 2\beta x,  \label{eq:p^2}
\end{equation}
where $C$ is some integration constant. Second, from (\ref{eq:p^2}), we deduce the equation
\begin{equation}
  dt = \frac{dx}{\sqrt{a + bx + cx^2}}, \qquad a = C + \frac{\alpha^2}{\lambda}, \qquad b = 2\beta, \qquad
  c = C \lambda,
\end{equation}
which can be easily integrated. For such a purpose, we use either Eq.~(2.02.10) of Ref.~\cite{gradshteyn} whenever $c=0$ or Eq.~(2.261) of the same and the sign of the discrimant $\Delta = 4ac - b^2$ whenever $c \ne 0$. The result $t = t(x)$ can then be inverted to yield $x = x(t)$.\par
%
%--------------------------------------------------------------------------------------------------------------------------
%
To provide some physically-relevant results, it is worth observing that the value of the integration constant $C$ is directly related to the energy $E$ of the system. The latter is indeed given by
\begin{equation}
  E = \frac{1}{2} \frac{1}{1 + \lambda x^2} (\dot{x}^2 + \alpha^2 x^2 - 2\beta x),
\end{equation}
which, when inserting (\ref{eq:p^2}), becomes
\begin{equation}
  E = \frac{1}{2} C + \frac{\alpha^2}{2\lambda} \qquad \mbox{or} \qquad C = 2E - \frac{\alpha^2}{\lambda}.
  \label{eq:E-C}
\end{equation}
The restrictions on the constants $C$, $c$, and $\Delta$ for each energy domain are presented in Table~I.\par
%
%---------------------------------------------------------------------------------------------------------------------
%
\begin{table}[h!]

\caption{Restrictions on $C$, $c$, and $\Delta$ corresponding to possible values of $\lambda$ and $E$ for the first generalized nonlinear oscillator.}

\begin{center}
\begin{tabular}{lcccl}
  \hline\hline\\[-0.2cm]
  $\lambda$ & $E$ & $C$ & $c$ & $\Delta$ \\[0.2cm]
  \hline\\[-0.2cm]
  $\lambda > 0$ & $V_{\rm min} < E < V(+\infty)$ & $- \frac{1}{2\lambda}\left(\alpha^2 + \sqrt{\alpha^4
      + 4\lambda \beta^2}\right) < C < 0$ & $c < 0$ & $\Delta < 0$ \\[0.2cm]
  $\lambda > 0$ & $E = V(+\infty)$ & $C = 0$ & $c = 0$ & ---  \\[0.2cm]
  $\lambda > 0$ & $V(+\infty) < E < V_{\rm max}$ & $0 < C < \frac{1}{2\lambda}\left(- \alpha^2 +
      \sqrt{\alpha^4 + 4\lambda \beta^2}\right)$ & $c > 0$ & $\Delta < 0$ \\[0.2cm]
  $\lambda > 0$ & $E = V_{\rm max}$ & $C = \frac{1}{2\lambda}\left(- \alpha^2 + \sqrt{\alpha^4 + 4\lambda 
      \beta^2}\right)$ & $c > 0$ & $\Delta = 0$ \\[0.2cm]
  $\lambda > 0$ & $V_{\rm max} < E < +\infty$ & $\frac{1}{2\lambda}\left(- \alpha^2 + \sqrt{\alpha^4 + 
      4\lambda \beta^2}\right) < C < +\infty$ & $c > 0$ & $\Delta > 0$ \\[0.2cm]
  $\lambda < 0$ & $V_{\rm min} < E < +\infty$ & $\frac{1}{2|\lambda|}\left(\alpha^2 + \sqrt{\alpha^4
      - 4|\lambda| \beta^2}\right) < C < +\infty$ & $c < 0$ & $\Delta < 0$ \\[0.2cm] 
  \hline \hline
\end{tabular}
\end{center}

\end{table}
\par
%
%----------------------------------------------------------------------------------------------------------------------
%
The solutions $x(t)$ so obtained are listed in Table~II, together with the values of the parameters $A$, $B$, $\phi$ that have been introduced. The (whenever relevant) `frequencies' $\omega$ are given in Table~III \cite{footnote}.\par
%
%------------------------------------------------------------------------------------------------------------------
%
\begin{table}[h!]

\caption{Solutions $x(t)$ of the Euler-Lagrange equation for the first generalized nonlinear oscillator.}

\begin{center}
\begin{tabular}{lccccl}
  \hline\hline\\[-0.2cm]
  $\lambda$ & $E$ & $x(t)$ & $A$ & $B$ & $\phi$ \\[0.2cm]
  \hline\\[-0.2cm]
  $\lambda > 0$ & $V_{\rm min} < E < V(+\infty)$ & $A \sin(\omega t + \phi) + B$ & $\frac{1}{\omega^2}
      \sqrt{\frac{- \omega^4 + \alpha^2 \omega^2 + \lambda \beta^2}{\lambda}}$ & $\frac{\beta}{\omega^2}$
      & $[0, 2\pi)$ \\[0.2cm]
  & & $B-A \le x \le B+A$ & & & \\[0.2cm]
  $\lambda > 0$ & $E = V(+\infty)$ & $(At + \phi)^2 + B$ & $\sqrt{\frac{\beta}{2}}$ & $- \frac{\alpha^2}
      {2\lambda \beta}$ & $\R$   \\[0.2cm]
  & & $B \le x < +\infty$ & & & \\[0.2cm]
  $\lambda > 0$ & $V(+\infty) < E < V_{\rm max}$ & $A \cosh(\omega t + \phi) + B$ & $\frac{1}
      {\omega^2} \sqrt{\frac{- \omega^4 - \alpha^2 \omega^2 + \lambda \beta^2}{\lambda}}$ & $-\frac{\beta}
      {\omega^2}$ & $\R$ \\[0.2cm]
  & & $A+B \le x < +\infty$ & & & \\[0.2cm]
  & & $A \cosh(\omega t + \phi) + B$ & $-\frac{1}{\omega^2} \sqrt{\frac{- \omega^4 - \alpha^2 \omega^2 +
      \lambda \beta^2}{\lambda}}$ & $-\frac{\beta}{\omega^2}$ & $\R$ \\[0.2cm]
  & & $-\infty < x \le A+B$ & & & \\[0.2cm]
  $\lambda > 0$ & $E = V_{\rm max}$ & $A e^{\omega t} + B$ & $\frac{1}{2\omega^2} e^{\phi}$ & $- \frac
      {\beta}{\omega^2}$ & $\R$ \\[0.2cm]
  & & $A+B \le x < +\infty$ & & & \\[0.2cm]
  & & $A e^{\omega t} + B$ & $-\frac{1}{2\omega^2} e^{\phi}$ & $- \frac
      {\beta}{\omega^2}$ & $\R$ \\[0.2cm]
  & & $-\infty < x \le A+B$ & & & \\[0.2cm]
  $\lambda > 0$ & $V_{\rm max} < E < +\infty$ & $A \sinh(\omega t + \phi) + B$ & $\pm \frac{1}
      {\omega^2} \sqrt{\frac{\omega^4 + \alpha^2 \omega^2 - \lambda \beta^2}{\lambda}}$ & $- \frac
      {\beta}{\omega^2}$ & $\R$ \\[0.2cm]
  & & $-\infty < x < +\infty$ & & & \\[0.2cm]
  $\lambda < 0$ & $V_{\rm min} < E < +\infty$ & $A \sin(\omega t + \phi) + B$ & $\frac{1}{\omega^2}
      \sqrt{\frac{\omega^4 - \alpha^2 \omega^2 + |\lambda| \beta^2}{|\lambda|}}$ & $\frac{\beta}
      {\omega^2}$ & $\R$ \\[0.2cm] 
  & & $B-A \le x \le B+A$ & & & \\[0.2cm]
  \hline \hline
\end{tabular}
\end{center}

\end{table}
\par
%
%----------------------------------------------------------------------------------------------------------------------
%
\begin{table}[h!]

\caption{`Frequency' $\omega$ in terms of the parameters of the Euler-Lagrange equation solutions obtained for the first generalized nonlinear oscillator.}

\begin{center}
\begin{tabular}{lcl}
  \hline\hline\\[-0.2cm]
  $\lambda$ & $E$ & $\omega^2 = |\lambda C|$ \\[0.2cm]
  \hline\\[-0.2cm]
  $\lambda > 0$ & $V_{\rm min} < E < V(+\infty)$ & $\frac{\alpha^2}{1 + \lambda(A^2-B^2)} =
      \frac{\alpha^2 + \sqrt{\alpha^4 + 4\lambda \beta^2 (1 + \lambda A^2)}}{2 (1 + \lambda A^2)}$ 
      \\[0.2cm]
  $\lambda > 0$ & $V(+\infty) < E < V_{\rm max}$ & $-\frac{\alpha^2}{1 + \lambda(A^2-B^2)} =
      \frac{- \alpha^2 + \sqrt{\alpha^4 + 4\lambda \beta^2 (1 + \lambda A^2)}}{2 (1 + \lambda A^2)}$ 
      \\[0.2cm]
  $\lambda > 0$ & $E = V_{\rm max}$ & $- \frac{\alpha^2}{1 - \lambda B^2} = \frac{1}{2} \left(- \alpha^2
      + \sqrt{\alpha^4 + 4\lambda \beta^2}\right)$ \\[0.2cm]
  $\lambda > 0$ & $V_{\rm max} < E < +\infty$ & $- \frac{\alpha^2}{1 - \lambda(A^2+B^2)} =
      \frac{- \alpha^2 + \sqrt{\alpha^4 + 4\lambda \beta^2 (1 - \lambda A^2)}}{2 (1 - \lambda A^2)}$ if 
      $\lambda B^2 > \lambda A^2 - 1$\\[0.2cm]
   & & $\hphantom{- \frac{\alpha^2}{1 - \lambda(A^2+B^2)}} =
      \frac{- \alpha^2 - \sqrt{\alpha^4 + 4\lambda \beta^2 (1 - \lambda A^2)}}{2 (1 - \lambda A^2)}$ if 
      $\lambda B^2 < \lambda A^2 - 1$\\[0.2cm]
  $\lambda < 0$ & $V_{\rm min} < E < +\infty$ & $\frac{\alpha^2}{1 - |\lambda|(A^2-B^2)} =
      \frac{\alpha^2 + \sqrt{\alpha^4 - 4|\lambda| \beta^2 (1 - |\lambda| A^2)}}{2 (1 - |\lambda| A^2)}$ 
      \\[0.2cm] 
  \hline \hline
\end{tabular}
\end{center}

\end{table}
\par
%
%----------------------------------------------------------------------------------------------------------------------
%
The new parameter $\beta$ in the potential is responsible for a new additive constant $B$ in the solutions, which also appears in $\omega$ and makes the dependence of the latter more complicated than the original one. Apart from this, the sine-dependent bounded solutions and the hyperbolic-sine-dependent  unbounded ones are rather similar to those previously obtained, to which they go over in the $\beta \to 0$ limit. In contrast, the hyperbolic-cosine-dependent unbounded solutions constitute a new type of solutions, with no counterpart in the $\beta \to 0$ limit. It is worth observing too that due to the presence of a bump in the potential, for each energy there are actually two solutions: one on the right extending over the interval $(B + |A|, +\infty)$, and one on the left, spreading over the interval $(-\infty, B - |A|)$. Finally, it may be noted that the limiting unbounded solution, which was linear, has been replaced by quadratic and exponential ones.\par
%
%===================================================================
% 
\section{SECOND GENERALIZED NONLINEAR OSCILLATOR}

\setcounter{equation}{0}

\subsection{The potential}

Let us next consider the potential
\begin{equation}
  V(x) = \frac{1}{2} \frac{\alpha^2 x^2 - 2\beta x \sqrt{1 + \lambda x^2}}{1 + \lambda x^2},
\end{equation}
depending on an additional parameter $\beta$ and yielding the Mathews and Lakshmanan potential for $\beta \to 0$ again. The two cases between which we distinguish according to the sign of $\lambda$ are now:
\begin{eqnarray}
  & ({\rm i}) \; &  \lambda < 0, \quad - \frac{1}{\sqrt{|\lambda|}} < x < \frac{1}{\sqrt{|\lambda|}}, \quad 
       \beta > 0;  \\
  & ({\rm ii}) \; &  \lambda > 0, \quad -\infty < x < \infty, \quad  0 < \beta < \frac{\alpha^2}{\sqrt{\lambda}}.
\end{eqnarray}
\par
%
%----------------------------------------------------------------------------------------------------------------------
% 
In the first case, the potential is rather similar to the previous one as it is a well with boundless walls at $x = \pm 1/\sqrt{|\lambda|}$, two zeros at $x=0$, $x = 2\beta/\sqrt{\alpha^4 + 4|\lambda| \beta^2}$, and a minimum in between,
\begin{equation}
  V_{\rm min} = V(x_{\rm min}) = - \frac{\beta^2}{2\alpha^2} \qquad \mbox{at} \qquad x_{\rm min} = 
  \frac{\beta}{\sqrt{\alpha^4 + |\lambda|\beta^2}}.
\end{equation}
In the second case, the potential looks rather different from the previous one. There is now no maximum, but the potential goes to different limits when $x \to \pm \infty$, namely $V(\pm \infty) = (\alpha^2 \mp 2\beta \sqrt{\lambda})/(2\lambda)$. Hence $V(- \infty)$ is always positive, whereas $V(+\infty)$ is positive, null, or negative according to whether $0 < \beta < \alpha^2/(2\sqrt{\lambda})$, $\beta = \alpha^2/(2\sqrt{\lambda})$, or $\alpha^2/(2\sqrt{\lambda}) < \beta < \alpha^2/\sqrt{\lambda}$. As a consequence, apart from a first zero at $x=0$, there is a second zero at $x = 2\beta/\sqrt{\alpha^4 - 4\lambda \beta^2}$, or at $x \to +\infty$, or no second zero at all, respectively. The presence or absence of a second zero is, however, not much relevant, because in all cases there is a minimum
\begin{equation}
  V_{\rm min} = V(x_{\rm min}) = - \frac{\beta^2}{2\alpha^2} \qquad \mbox{at} \qquad x_{\rm min} = 
  \frac{\beta}{\sqrt{\alpha^4 - \lambda\beta^2}}.
\end{equation}
\par
%
%----------------------------------------------------------------------------------------------------------------
%
\begin{figure}[h]
\begin{center}
\includegraphics{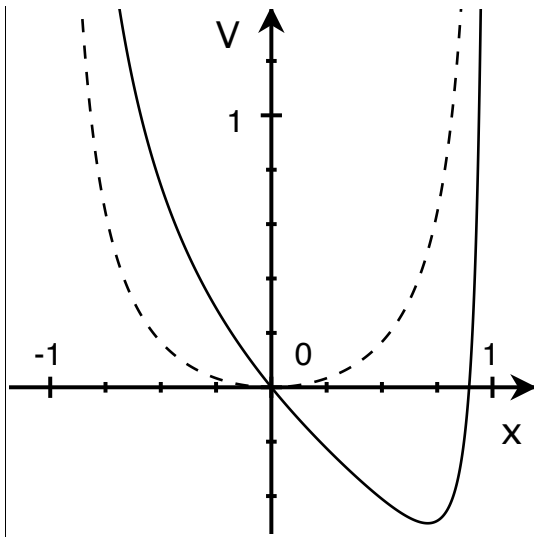}
\caption{Plot of $V(x)= \tfrac{1}{2} (\alpha^2 x^2 - 2\beta x \sqrt{1 + \lambda x^2})/(1 + \lambda x^2)$, $\alpha = -\lambda = 1$, as a function of $x$, for $\beta = 1$ (solid line) and $\beta = 0$ (dashed line).}
\end{center}
\end{figure}
\par
%
%-------------------------------------------------------------------------------------------------------------------------------
%
\begin{figure}[h]
\begin{center}
\includegraphics{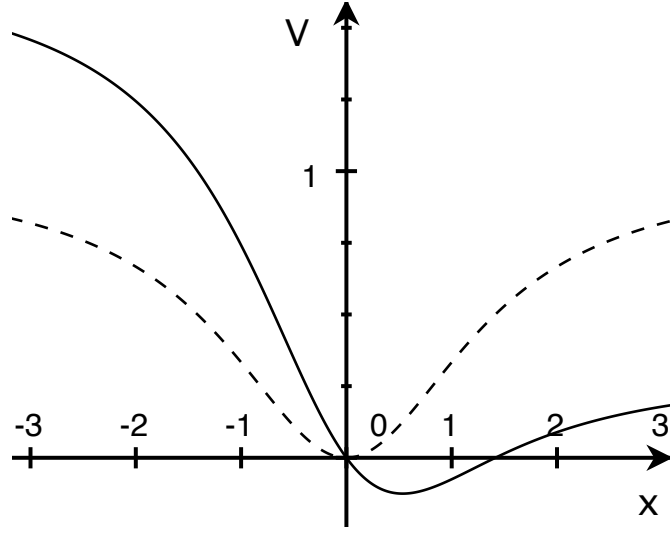}
\caption{Plot of $V(x)= \tfrac{1}{2} (\alpha^2 x^2 - 2\beta x \sqrt{1 + \lambda x^2})/(1 + \lambda x^2)$, $\alpha = 1$, $\lambda = 0.5$, as a function of $x$, for $\beta = 0.5$ (solid line) and $\beta = 0$ (dashed line).}
\end{center}
\end{figure}
\par
%
%-------------------------------------------------------------------------------------------------------------------------------
%
In Figs.~3 and 4, some examples are displayed and compared with the $\beta = 0$ case. Bounded solutions will again be present for all energies if $\lambda < 0$ and for small energies if $\lambda > 0$. In the latter case, as before we will observe a splitting of unbounded solutions, this time between energies below or above $V(-\infty)$, and, as a result, the existence of two limiting unbounded motions at $E = V(+\infty)$ and $E = V(-\infty)$, respectively.\par
%
%+++++++++++++++++++++++++++++++++++++++++++++++++++++++++++++++++
%
\subsection{Euler-Lagrange equation}

The Euler-Lagrange equation now reads
\begin{equation}
  (1 + \lambda x^2) \ddot{x} - \lambda x \dot{x}^2 + \alpha^2 x - \beta \sqrt{1 + \lambda x^2} = 0.
  \label{eq:E-L-2}
\end{equation}
On setting $\dot{x} = p(x)$, we obtain the first-order equation
\begin{equation}
  (1 + \lambda x^2) \frac{dp^2}{dx} - 2\lambda x p^2 + 2\alpha^2 x - 2\beta \sqrt{1 + \lambda x^2} = 0,
\end{equation}
whose general solution is
\begin{equation}
  p^2(x) = C (1 + \lambda x^2) + \frac{\alpha^2}{\lambda} + 2\beta x \sqrt{1 + \lambda x^2}.
  \label{eq:p^2-bis}
\end{equation}
Here the integration constant $C$ is related to the energy $E$ through Eq.~(\ref{eq:E-C}) again. In Table IV, we list the restrictions on $C$ corresponding to possible values of $\lambda$ and $E$.\par
%
%--------------------------------------------------------------------------------------------------------------------------
%
\begin{table}[h!]

\caption{Restrictions on $C$ corresponding to possible values of $\lambda$ and $E$ for the second generalized nonlinear oscillator.}

\begin{center}
\begin{tabular}{lcl}
  \hline\hline\\[-0.2cm]
  $\lambda$ & $E$ & $C$ \\[0.2cm]
  \hline\\[-0.2cm]
  $\lambda > 0$ & $V_{\rm min} < E < V(+\infty)$ & $- \frac{\alpha^2}{\lambda} - \frac{\beta^2}{\alpha^2}
      < C < - \frac{2\beta}{\sqrt{\lambda}}$ \\[0.2cm]
  $\lambda > 0$ & $E = V(+\infty)$ & $C = - \frac{2\beta}{\sqrt{\lambda}}$ \\[0.2cm]
  $\lambda > 0$ & $V(+\infty) < E < V(-\infty)$ & $- \frac{2\beta}{\sqrt{\lambda}} < C < \frac{2\beta}
      {\sqrt{\lambda}}$ \\[0.2cm]
  $\lambda > 0$ & $E = V(-\infty)$ & $C = \frac{2\beta}{\sqrt{\lambda}}$ \\[0.2cm]
  $\lambda > 0$ & $V(-\infty) < E < +\infty$ & $\frac{2\beta}{\sqrt{\lambda}} < C < +\infty$ \\[0.2cm]
  $\lambda < 0$ & $V_{\rm min} < E < +\infty$ & $\frac{\alpha^2}{|\lambda|} - \frac{\beta^2}{\alpha^2} < C 
      < +\infty$ \\[0.2cm] 
  \hline \hline
\end{tabular}
\end{center}

\end{table}
\par
%
%----------------------------------------------------------------------------------------------------------------------
%
In the next step, Eq.~(\ref{eq:p^2-bis}) is transformed into
\begin{equation}
  dt = \frac{dx}{\left(C + \frac{\alpha^2}{\lambda} + 2\beta x \sqrt{1 + \lambda x^2} + \lambda C x^2
  \right)^{1/2}},  \label{eq:diff-eqn}
\end{equation}
which can be integrated using several changes of variable. Since the latter are different for $\lambda > 0$ and $\lambda < 0$, we shall proceed to consider both cases successively.\par
%
%xxxxxxxxxxxxxxxxxxxxxxxxxxxxxxxxxxxxxxxxxxxxxxxxxxxxxxxxxxxxxxxxxxx
%
\subsubsection{\boldmath Solutions of the Euler-Lagrange equation for $\lambda > 0$}

By writing
\begin{equation}
  u = \frac{\sqrt{\lambda}\, x}{\sqrt{1 + \lambda x^2}}, \qquad v = u - 1, \qquad w = u + 1,
\end{equation}
the solutions $t = t(x)$ of (\ref{eq:diff-eqn}) can be split into the sum of two integrals
\begin{equation}
  t + K = I_1 + I_2,  \label{eq:t}
\end{equation}
where $K$ is some integration constant and
\begin{equation}
  I_1 = - \frac{1}{2} \int \frac{dv}{v \sqrt{a + bv + cv^2}}, \quad a = \lambda C + 2\beta \sqrt{\lambda},
  \quad b = - 2(\alpha^2 - \beta \sqrt{\lambda}), \quad c = - \alpha^2,
\end{equation}
\begin{equation}
  I_2 = \frac{1}{2} \int \frac{dw}{w \sqrt{a' + b'w + c' w^2}}, \quad a' = \lambda C - 2\beta \sqrt{\lambda},
  \quad b' = 2(\alpha^2 + \beta \sqrt{\lambda}), \quad c' = - \alpha^2.
\end{equation}
Both of these integrals can be performed by using Eq.~(2.266) of Ref.~\cite{gradshteyn}. Since the corresponding discriminants $\Delta = 4ac - b^2$ and $\Delta' = 4a'c' - b^{\prime 2}$ are such that $\Delta = \Delta' = - 8 \alpha^2 \lambda (E - V_{\rm min}) < 0$, the results will only depend on the sign of $a$ or $a'$. These parameters can be rewritten as $a = 2\lambda [E - V(+\infty)]$ and $a' = 2 \lambda [E - V(-\infty)]$, hence we may write
\begin{equation}
  I_1 = \begin{cases}
     I_{11} & \text{if $E < V(+\infty)$}, \\
     I_{12} & \text{if $E = V(+\infty)$}, \\ 
     I_{13} & \text{if $E > V(+\infty)$},    
  \end{cases}
\end{equation}
and
\begin{equation}
  I_2 = \begin{cases}
     I_{21} & \text{if $E < V(-\infty)$}, \\
     I_{22} & \text{if $E = V(-\infty)$}, \\ 
     I_{23} & \text{if $E > V(-\infty)$}.    
  \end{cases}
\end{equation}
\par
%
%--------------------------------------------------------------------------------------------------------------------
%
After some straightforward calculations, the $I_{ij}$'s can be expressed in terms of $x$ and the parameters as
\begin{equation}
\begin{split}
  I_{11} &= \frac{1}{2 \sqrt{2\lambda [V(+\infty) - E]}} \\
  & \quad \times \arcsin \frac{-2\lambda [V(+\infty) - E] \sqrt{1+\lambda x^2}(\sqrt{\lambda} x + 
       \sqrt{1+\lambda x^2}) + \alpha^2 - \beta \sqrt{\lambda}}{\alpha \sqrt{2\lambda (E - V_{\rm min})}}, \\
  I_{12} &= \frac{1}{2(\alpha^2 - \beta \sqrt{\lambda})} \left(\sqrt{\lambda} x + \sqrt{1+\lambda x^2}\right) 
       \\
  & \quad \times\left(\alpha^2 - 2\beta \sqrt{\lambda} + 2\beta \lambda x \sqrt{1+\lambda x^2} 
       - 2\beta \lambda^{3/2} x^2\right)^{1/2}, \\
  I_{13} &= \frac{1}{2 \sqrt{2\lambda [E - V(+\infty)]}} \\
  & \quad \times \ln \Bigl\{-4\lambda [E - V(+\infty)] 
       \sqrt{1+\lambda x^2}\left(\sqrt{\lambda} x + \sqrt{1+\lambda x^2}\right) 
       - 2(\alpha^2 - \beta \sqrt{\lambda}) \\
  & \quad - 4\lambda  \sqrt{E - V(+\infty)} \left(\sqrt{\lambda} x + \sqrt{1+\lambda x^2}\right) \\
  & \quad \times \Bigl[E + \beta x 
       \sqrt{1+\lambda x^2} + \Bigl(E - \frac{\alpha^2}{2\lambda}\Bigr) \lambda x^2\Bigr]^{1/2}\Bigr\}, \\
  I_{21} &= \frac{1}{2 \sqrt{2\lambda [V(-\infty) - E]}} \\
  & \quad \times \arcsin \frac{-2\lambda [V(-\infty) - E] \sqrt{1+\lambda x^2}(- \sqrt{\lambda} x + 
       \sqrt{1+\lambda x^2}) + \alpha^2 + \beta \sqrt{\lambda}}{\alpha \sqrt{2\lambda (E - V_{\rm min})}}, \\ 
  I_{22} &= \frac{1}{2(\alpha^2 + \beta \sqrt{\lambda})} \left(\sqrt{\lambda} x - \sqrt{1+\lambda x^2}\right) 
       \\
  & \quad \times\left(\alpha^2 + 2\beta \sqrt{\lambda} + 2\beta \lambda x \sqrt{1+\lambda x^2} 
       + 2\beta \lambda^{3/2} x^2\right)^{1/2}, \\
  I_{23} &= - \frac{1}{2 \sqrt{2\lambda [E - V(-\infty)]}} \\
  & \quad \times \ln \Bigl\{4\lambda [E - V(-\infty)] 
       \sqrt{1+\lambda x^2}\left(- \sqrt{\lambda} x + \sqrt{1+\lambda x^2}) + 2(\alpha^2 + \beta 
       \sqrt{\lambda}\right) \\
  & \quad + 4\lambda  \sqrt{E - V(-\infty)} \left(- \sqrt{\lambda} x + \sqrt{1+\lambda x^2}\right) \\
  & \quad \times \Bigl[E + \beta x 
       \sqrt{1+\lambda x^2} + \Bigl(E - \frac{\alpha^2}{2\lambda}\Bigr) \lambda x^2\Bigr]^{1/2}\Bigr\}.               
\end{split}
\end{equation}
\par
%
%-------------------------------------------------------------------------------------------------------------------------
%
By combining all the results, we finally obtain the solution as
\begin{equation}
  t+K = \begin{cases}
     I_{11} + I_{21} & \text{if $V_{\rm min} < E < V(+\infty)$}, \\
     I_{12} + I_{21} & \text{if $E = V(+\infty)$}, \\ 
     I_{13} + I_{21} & \text{if $V(+\infty) < E < V(-\infty)$}, \\
     I_{13} + I_{22} & \text{if $E = V(-\infty)$}, \\
     I_{13} + I_{23} & \text{if $V(-\infty) < E < + \infty$}.    
  \end{cases}
\end{equation}
This provides us with an implicit solution $t=t(x)$ of the Euler-Lagrange equation (\ref{eq:E-L-2}), since its inverse $x=x(t)$ cannot be written in closed form.\par
%
%xxxxxxxxxxxxxxxxxxxxxxxxxxxxxxxxxxxxxxxxxxxxxxxxxxxxxxxxxxxxxxxxxxxxxxxxxxxxxxxxxxxx
%
\subsubsection{\boldmath Solutions of the Euler-Lagrange equation for $\lambda < 0$}

In the present case, let us make the changes of variable
\begin{equation}
  u = \frac{\sqrt{|\lambda|}\, x}{\sqrt{1 - |\lambda| x^2}}, \qquad v = - \frac{2\beta u + \sqrt{|\lambda|}\, C
  + \sqrt{|\lambda|\, C^2 + 4\beta^2}}{2\beta u + \sqrt{|\lambda|}\, C - \sqrt{|\lambda|\, C^2 + 4\beta^2}},
\end{equation}
yielding a splitting of $t=t(x)$ of type (\ref{eq:t}), where $I_1$ and $I_2$ are now given by
\begin{equation}
  I_1 = L \int \frac{v dv}{(v^2 + p) \sqrt{v^2 - q}}, \qquad I_2 = L \int \frac{dv}{(v^2 + p) \sqrt{v^2 - q}}.
  \label{eq:I_1-I_2}
\end{equation}
Here $L$, $p$, and $q$ are three constants, defined by
\begin{equation}
\begin{split}
  L &= 2\sqrt{2}\, \beta^2 (|\lambda|\, C^2 + 4\beta^2)^{-1/4} \left(\sqrt{|\lambda|\, C^2 + 4\beta^2}
       - \sqrt{|\lambda|}\, C \right)^{-1} \\
  & \quad \times \left[\sqrt{|\lambda|} (\alpha^2 C + 2\beta^2) - \alpha^2 \sqrt{|\lambda|\, C^2 + 4\beta^2}
       \right]^{-1/2},
\end{split}
\end{equation}
\begin{equation}
  p = \frac{1}{2\beta^2} \left(|\lambda|\, C^2 + 2\beta^2 + \sqrt{|\lambda|}\, C \sqrt{|\lambda|\, C^2 + 
  4\beta^2}\right) > 0,
\end{equation}
\begin{equation}
  q = \frac{\sqrt{|\lambda|} (\alpha^2 C + 2\beta^2) + \alpha^2 \sqrt{|\lambda|\, C^2 + 4\beta^2}}
  {\sqrt{|\lambda|} (\alpha^2 C + 2\beta^2) - \alpha^2 \sqrt{|\lambda|\, C^2 + 4\beta^2}} > 0.
\end{equation}
\par
%
%---------------------------------------------------------------------------------------------------------------------
%
The integrals $I_1$ and $I_2$ of Eq.~(\ref{eq:I_1-I_2}) can be finally reduced to some elementary integrals by performing some additional changes of variable, namely
\begin{equation}
  w = \sqrt{\frac{v^2 - q}{p+q}} \qquad \mbox{and} \qquad z = \sqrt{\frac{p+q}{p}} \frac{v}{\sqrt{v^2 - q}}.
\end{equation}
The results read
\begin{equation}
  I_1 = \frac{L}{\sqrt{p+q}} \arctan w, \qquad I_2 = \frac{L}{\sqrt{p(p+q)}} \arctanh z,
\end{equation}
where
\begin{equation}
\begin{split}
  p+q &= \frac{\sqrt{|\lambda|\, C^2 + 4\beta^2}}{2\beta^2 (|\lambda| \alpha^2 C + |\lambda| \beta^2
       - \alpha^4)} \\
  & \quad \times \left[|\lambda| (\alpha^2 C + \beta^2) \left(\sqrt{|\lambda|}\, C + \sqrt{|\lambda|\, C^2 +
       4\beta^2}\right) + 2 \sqrt{|\lambda|}\, \alpha^2 \beta^2\right].
\end{split}
\end{equation}
\par
%
%--------------------------------------------------------------------------------------------------------------------
%
To complete the results, it is worth mentioning the explicit form of $v$ in terms of $x$,
\begin{equation}
\begin{split}
  v &= 2\beta \left(\beta + \sqrt{|\lambda| C^2 + 4\beta^2} \sqrt{|\lambda|} x \sqrt{1 - |\lambda| x^2}\right)
      \Bigl[|\lambda| C^2 + 2\beta^2 \\
  & \quad - \sqrt{|\lambda|} C \sqrt{|\lambda| C^2 + 4\beta^2} - \sqrt{|\lambda| C^2 + 4\beta^2} 
      \left(- \sqrt{|\lambda|} C + \sqrt{|\lambda| C^2 + 4\beta^2}\right) |\lambda| x^2\Bigr]^{-1},
\end{split}
\end{equation}
from which those of $w$ and $z$ could be easily obtained.\par
%
%=========================================================================
%
\section{CONCLUSION}

In the present paper, we have considered two generalizations of the Mathews and Lakshmanan nonlinear oscillator with the same kinetic energy term, but an extra term in the potential. We have derived the explicit (resp.\ implicit) solutions of the Euler-Lagrange equation corresponding to the first (resp.\ second) potential and we have shown an increased richness in their behaviour due to the more involved form of the potentials. It is worth mentioning that the quantum version of these generalized nonlinear oscillators also proves exactly solvable \cite{midya}.\par
%
%---------------------------------------------------------------------------------------------------------------------
%
An interesting open question for future work is whether the generalized Mathews and Lakshmanan nonlinear oscillators considered here could be extended to two dimensions in a way similar to (or different from) what has been done in the standard case \cite{carinena04b, carinena07b, carinena07c}.\par
%
%============================================================================
%


\begin{thebibliography}{99}
\bibitem{footnote}
Strictly speaking, $\omega$ may be considered as a frequency only in the case of bounded motions.
\end{thebibliography}
\end{document}